\documentclass[conference]{IEEEtran}
\IEEEoverridecommandlockouts
\usepackage{listings}
\usepackage{cite}
\usepackage{amsmath,amssymb,amsfonts}
\usepackage[utf8]{inputenc}
\usepackage{graphicx}
\usepackage{tikz}
\usetikzlibrary{shapes.geometric, arrows, positioning}
\usepackage{textcomp}
\usepackage{xcolor}
\usepackage[linesnumbered,ruled,vlined]{algorithm2e}
\usepackage{caption}
\usepackage{booktabs}

\usepackage[linesnumbered,ruled,vlined]{algorithm2e}

\usepackage{xcolor}
\definecolor{mycommentcolor}{rgb}{0.0, 0.5, 0.0}
\definecolor{mykeywordcolor}{rgb}{0.2, 0.2, 0.6}
\definecolor{myvarcolor}{rgb}{0.5, 0.0, 0.0}

\SetCommentSty{mycommentstyle}

\SetKwSty{textbf}{\color{mykeywordcolor}}{}

\def\BibTeX{{\rm B\kern-.05em{\sc i\kern-.025em b}\kern-.08em
    T\kern-.1667em\lower.7ex\hbox{E}\kern-.125emX}}

\begin{document}

\title{Toward a Lightweight, Scalable, and Parallel \\Secure Encryption Engine}

\author{
\begin{tabular}{cc}
\begin{minipage}{0.45\linewidth}
\centering
Rasha Karakchi \\
Dept. of Computer Science and Engineering \\
University of South Carolina \\
Columbia, USA \\
karakchi@cec.sc.edu
\end{minipage}
&
\begin{minipage}{0.45\linewidth}
\centering
Rye Stahle-Smith \\
Dept. of Computer Science and Engineering \\
University of South Carolina \\
Columbia, USA \\
rye@email.sc.edu
\end{minipage}
\\[50pt]  
\begin{minipage}{0.45\linewidth}
\centering
Nishant Chinnasami\\
Dept. of Computer Science and Engineering \\
University of South Carolina \\
Columbia, USA \\
nishantc@email.sc.edu
\end{minipage}
&
\begin{minipage}{0.45\linewidth}
\centering
Tiffany Yu \\
Dept. of Computer Science and Engineering \\
University of South Carolina \\
Columbia, USA \\
tyu@email.sc.edu
\end{minipage}
\end{tabular}
}

\maketitle

\begin{abstract}
The exponential growth of applications of the Internet of Things (IoT) has intensified the demand for efficient, high-throughput, and energy-efficient data processing at the edge. Conventional CPU-centric encryption methods suffer from performance bottlenecks and excessive data movement, especially in latency-sensitive and resource-constrained environments. In this paper, we present SPiME, a lightweight, scalable, and FPGA-compatible Secure Processor in Memory Encryption architecture that integrates the Advanced Encryption Standard (AES-128) directly into a Processing-in-Memory (PiM) framework. SPiME is designed as a modular array of parallel PiM units, each combining an AES core with a minimal control unit to enable distributed in-place encryption with minimal overhead.

The architecture is fully implemented in Verilog and tested on multiple AMD UltraScale and UltraScale+ FPGAs. Evaluational results show that SPiMe can scale beyond 4,000 parallel units while maintaining less than 5\% utilization of key FPGA resources on high-end devices. It delivers over 25 Gbps in sustained encryption throughput with predictable, low-latency performance. The design’s portability, configurability, and resource efficiency make it a compelling solution for secure edge computing, embedded cryptographic systems, and customizable hardware accelerators.
\end{abstract}

\begin{IEEEkeywords}
FPGA, Verilog, AES, Processor-in-Memory
\end{IEEEkeywords}

\section{Introduction}
The rise of Internet of Things (IoT) systems has significantly accelerated Big Data generation, raising urgent challenges in secure, real-time, and energy-efficient data processing \cite{yang2019processing}. IoT devices continuously produce sensitive data streams that must be encrypted and processed under tight bandwidth and energy constraints, exposing the limitations of conventional CPU-centric systems in terms of latency and data movement overhead.

Processing-in-Memory (PiM) has emerged as a promising alternative, enabling computation near data to reduce transfer overhead and improve performance \cite{yang2019processing}. While PiM has demonstrated benefits in domains such as pattern matching \cite{karakchi2023napoly, karakchi2017dynamically, karakchi2019overlay, karbowniczak2025optimizing}, genomics \cite{karakchi2016hls}, and AI inference \cite{karakchi2024developing}, securing data within PiM remains a critical challenge. Traditional memory hierarchies and centralized encryption methods fail to meet the urgent needs of modern IoT and edge computing environments \cite{jarvinen2008fully, xu2023pim}.

Security in PiM systems is particularly sensitive due to the risk of data exposure during transfer, susceptibility to side-channel attacks, and power leakage \cite{zhang2021energy}. The Advanced Encryption Standard (AES) is a widely adopted countermeasure, but software-based AES is computationally intensive and ill-suited for real-time applications \cite{mcevoy2006compact}. Embedding AES directly into PiM architectures mitigates these issues by protecting data in place, reducing latency, and minimizing the observable attack surface \cite{wang2021high, tiri2003securaes}.

In this work, we introduce \textbf{SPiME}, a scalable and lightweight AES-128 encryption system embedded within a PiM framework and implemented in Verilog. Each PiM unit integrates an \texttt{aes\_core}—comprising submodules \texttt{sub\_bytes}, \texttt{shift\_rows}, and \texttt{mix\_columns}—with a \texttt{pim\_controller} for key scheduling and I/O. The system supports throughput scaling via a parameterized \texttt{NUM\_PIMs} variable, enabling parallel encryption across distributed memory banks. SPiMe is designed for FPGA compatibility, modular reuse, and energy-efficient secure computation at the edge. It addresses critical performance and security requirements for PiM-based IoT platforms, providing a practical foundation for secure, high-throughput data processing.

\section{Related Work}
AES has been widely implemented in hardware to improve performance, energy efficiency, and area utilization~\cite{chodowiec2003very, deshpande2009fpga, priya2022fpga, zambreno2004exploring, chodowiec2002asic, good2005aes, zodpe2020efficient, hasija2023survey, deshpande2014efficient, borkar2011fpga, farooq2017comparative, zhang2018optimization, deshpande2015area, sunil2020implementation}. Prior designs have focused on optimizing resource-constrained implementations~\cite{satoh2001compact, mcevoy2006compact}, improving resistance to side-channel attacks through balanced logic~\cite{tiri2003securaes}, and achieving high throughput using pipelined and parallel structures~\cite{jarvinen2008fully, he2020parallel}. Other efforts are application-specific, such as targeting 5G networks~\cite{wang2021high} or employing approximate computing to save energy~\cite{zhang2021energy}, though these may compromise security. Xu et al.~\cite{xu2023pim} introduced a PiM-based AES integrated into DRAM, but their approach relies on custom memory technology, limiting portability and compatibility with FPGA platforms.

Chaves et al.~\cite{chaves2006reconfigurable} presented a polymorphic AES core that merges SubBytes and MixColumns operations to reduce resource usage and improve throughput. Iranfar et al.~\cite{inspintronic} developed a spintronic-based AES design for PiM, which offers strong resistance to power-based side-channel attacks through a symmetrical logic structure and uniform power profile. Liu et al.~\cite{liu2022aespim} proposed AESPIM for real-time video encryption, incorporating system-level enhancements such as data/user-level parallelism and QoS-aware scheduling to improve streaming performance. Instruction-level approaches~\cite{lee2010processor} leverage AES-specific ISA extensions or general-purpose operations like Pread and byte\_perm, achieving fast encryption on CPUs but relying on specialized hardware and typically supporting only non-feedback encryption modes.

In contrast, our work introduces a modular and reconfigurable FPGA-based AES architecture designed for Processing-in-Memory (PiM) systems. By integrating multiple AES-128 cores with lightweight controllers, our design enables parallel and pipelined encryption directly near memory. This significantly reduces data movement and memory access overhead, leading to improved throughput and energy efficiency. Unlike solutions that require custom memory or CPU instruction extensions, our architecture supports flexible deployment across FPGA platforms, making it well-suited for secure, scalable, and energy-aware applications in edge and embedded environments.

\begin{figure}[h]
\centering
\includegraphics[width=0.35\textwidth]{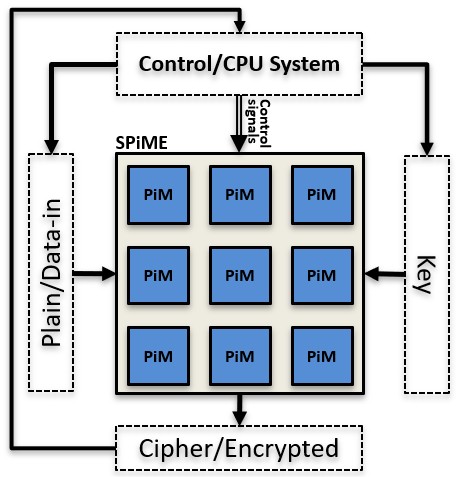} 
\caption{Proposed SPiME System.}
\label{fig:spimesys}
\end{figure}

\section{Design and Architectures}
Our proposed architecture SPiME (Secure Processor in Memory Encryption) is designed as an array of parallel multiple processor-in-memory (PiM) units where each PiM comprises a PiM controller and an AES core. The design is fully described in Verilog and features a modular organization that promotes scalability, reusability, and clarity. 

Figure \ref{fig:spimesys} illustrates the full SPiME architecture, composed of an array of multiple PiM units. Each PiM unit receives its plaintext input (data\_in) and encryption key from buffers managed by the CPU. The operation of the PiM units is directed by a set of control signals originating from the CPU. After processing, each PiM unit outputs the ciphertext to a corresponding output buffer. The experimental evaluation presented in this work focuses specifically on the SPiME component itself, excluding analysis of the resource usage or performance of the surrounding system. The followning subsections are description of SPiME and its components. 

\subsection{Top-Level System}
\begin{figure}[!h]
    \centering
    \includegraphics[width=0.45\textwidth]{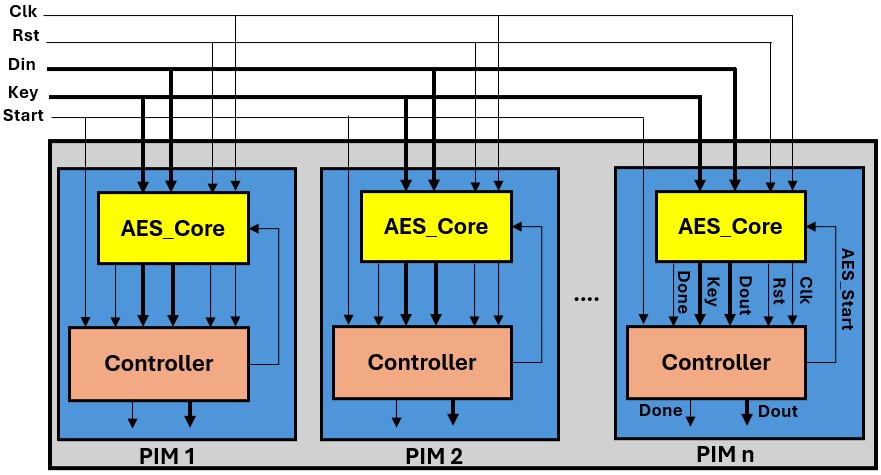} 
    \caption{SPiME top\_Level design consisting of an \textit{n} PiMs where each consists of AES\_Core and control unit}
    \label{fig:top_design}
\end{figure}

This is the main module that instantiates multiple PiM processing units, each consisting of a PiM\_Controller and an associated AES\_Core. It includes a parameter, NUM\_PiMs, which defines the number of parallel encryption units operating in the system. The top-level module coordinates global input/control signals such as system clock, reset, encryption start, and plaintext/key input. It gathers outputs from each processing unit and can be extended to include aggregation, output buffering, or interfacing to a larger memory subsystem. Figure \ref{fig:top_design} presents the design units of $n$ processor-in-memory units.

\subsection{PiM Controller} 
Each PiM\_Controller is responsible for orchestrating the AES encryption process for its assigned memory block. As shown in Algorithm \ref{alg:pim_controller}, it handles:

\textbf{Start/Done Handshaking} initiates the AES core when start is asserted, and monitors the done signal from the AES core.

\begin{algorithm}[!h]
\caption{PIM Controller State Machine}
\label{alg:pim_controller}
\scriptsize
\SetKwFunction{PIMController}{PIM\_Controller}
\SetKwProg{Fn}{Function}{:}{}
\Fn{\PIMController{clk, rst, start, data\_in, key, aes\_done, aes\_data\_out}}{

    \tcp{Initialize state}
    $state \gets \text{IDLE}$\;
    $aes\_start \gets 0$\;
    $done \gets 0$\;
    $data\_out \gets 0$\;

    \While{true}{
        \If{$rst = 1$}{
            $state \gets \text{IDLE}$\;
            $aes\_start \gets 0$\;
            $done \gets 0$\;
        }
        \Else{
            \Switch{$state$}{
                \Case{\text{IDLE}}{
                    \If{$start = 1$}{
                        $aes\_start \gets 1$\;
                        $state \gets \text{START\_AES}$\;
                    }
                }
                \Case{\text{START\_AES}}{
                    $aes\_start \gets 0$\;
                    $state \gets \text{WAIT\_AES}$\;
                }
                \Case{\text{WAIT\_AES}}{
                    \If{$aes\_done = 1$}{
                        $data\_out \gets aes\_data\_out$\;
                        $done \gets 1$\;
                        $state \gets \text{DONE}$\;
                    }
                }
                \Case{\text{DONE}}{
                    $done \gets 0$\;
                    $state \gets \text{IDLE}$\;
                }
            }
        }
    }
}
\end{algorithm}
\textbf{Data Handling} routes the input plaintext and key to the AES core and captures the result of the ciphertext after encryption.

\textbf{Control Flow Management} can support additional logic for data buffering, memory interfacing, or result distribution in a complete PIM setup. The controller acts as a lightweight scheduler that isolates control complexity from the AES datapath and makes the design modular.

\subsection{AES\_Core} It performs AES-128 encryption on a single 128-bit data block using a 128-bit key. It is pipelined and driven by a finite-state machine (FSM) that sequences through the AES rounds. In this design, we depend on the previous work of AES \cite{chodowiec2003very, deshpande2009fpga, priya2022fpga, zambreno2004exploring, chodowiec2002asic, good2005aes, zodpe2020efficient, hasija2023survey, deshpande2014efficient, borkar2011fpga, farooq2017comparative, zhang2018optimization, deshpande2015area, sunil2020implementation}. 
The \texttt{AES\_Core} is designed as a synchronous digital hardware module implemented in Verilog that controls the AES encryption process. This module coordinates the sequence of AES transformations applied to the input plaintext using a set of pre-computed round keys. 
\begin{itemize}
    \item \textbf{Inputs:}
    \begin{itemize}
        \item \texttt{clk}: System clock signal drives the sequential logic.
        \item \texttt{rst}: Active-high synchronous reset signal to initialize internal states.
        \item \texttt{start}: Signal to initiate the encryption process.
        \item \texttt{data\_in}: 128-bit plaintext input data.
        \item \texttt{key}: 128-bit AES key (used for key expansion).
        \item \texttt{round\_keys\_flat}: Concatenated 1408-bit bus containing all 11 round keys (each 128-bit).
    \end{itemize}
    \item \textbf{Outputs:}
    \begin{itemize}
        \item \texttt{data\_out}: 128-bit ciphertext output after encryption.
        \item \texttt{done}: Flag indicating completion of encryption.
    \end{itemize}
\end{itemize}

Upon initialization or reset, the module unpacks the \texttt{round\_keys\_flat} input into an array \texttt{round\_keys[0..10]}, each holding a 128-bit round key. The module includes a finite-state machine (FSM) with four states: \texttt{IDLE}, \texttt{INIT}, \texttt{ROUND}, and \texttt{FINAL}, which control the encryption flow:

 \textbf{IDLE State:} The module waits for the \texttt{start} signal. When asserted, it transitions to the \texttt{INIT} state. Outputs remain inactive during this state.
 
\textbf{INIT State:} Resets the round counter to zero and performs the initial AddRoundKey operation by XOR-ing the input data with \texttt{round\_keys[0]}, initializing the internal AES state. Then, it moves to the \texttt{ROUND} state.

\textbf{ROUND State:} Sequentially performs AES round transformations:\textit{SubBytes}, \textit{ShiftRows} and \textit{MixColumns}. 
The transformed state is XOR-ed with the next round key \texttt{round\_keys[round + 1]}, and the round counter is incremented. If the round counter reaches 9, the FSM transitions to the \texttt{FINAL} state; otherwise, it continues processing rounds.

\textbf{FINAL State:} Executes the final AES round, applying \textit{SubBytes} and \textit{ShiftRows} but skipping \textit{MixColumns}. The state is then XOR-ed with the last round key \texttt{round\_keys[10]}. The output \texttt{data\_out} is set to the encrypted data, \texttt{done} is asserted, and the FSM returns to \texttt{IDLE}. 

When the synchronous reset \texttt{rst} is asserted, all internal registers, including the state machine and round counter, are cleared. The FSM returns to the \texttt{IDLE} state, and the \texttt{done} signal is de-asserted. The AES\_Core is divided into the following submodules that represent the functional blocks of AES. Each submodule is designed to be synthesized and modular. Algorithm \ref{alg:aes_core} describes thoroughly the AES\_Core operation.

\begin{algorithm}[!h]
\caption{AES Core Algorithm}
\label{alg:aes_core}
\scriptsize
\SetKwFunction{AESCore}{AES\_Core}
\SetKwProg{Fn}{Function}{:}{}
\Fn{\AESCore{$clk, rst, start, data\_in, key, round\_keys\_flat$}}{
    \KwIn{$clk$, $rst$, $start$, $data\_in$, $key$, $round\_keys\_flat$}
    \KwOut{$done$, $data\_out$}

    \tcp{Unpack round keys from flattened input}
    \For{$i \gets 0$ \KwTo $10$}{
        $round\_keys[i] \gets round\_keys\_flat[i \cdot 128 +: 128]$\;
    }

    \tcp{Initialize FSM}
    $current\_state \gets IDLE$\;

    \tcp{FSM transition}
    \If{$rst$}{
        $current\_state \gets IDLE$\;
    }
    \Else{
        $current\_state \gets next\_state$\;
    }

    \tcp{FSM next state logic}
    \uIf{$current\_state = IDLE$}{
        $next\_state \gets start \ ?\ INIT : IDLE$\;
    }
    \uElseIf{$current\_state = INIT$}{
        $next\_state \gets ROUND$\;
    }
    \uElseIf{$current\_state = ROUND$}{
        $next\_state \gets (round = 9) \ ?\ FINAL : ROUND$\;
    }
    \uElseIf{$current\_state = FINAL$}{
        $next\_state \gets IDLE$\;
    }
    \Else{
        $next\_state \gets IDLE$\;
    }

    \tcp{Round logic}
    \If{$rst$}{
        $state \gets 0$\;
        $round \gets 0$\;
        $done \gets 0$\;
    }
    \Else{
        \uIf{$current\_state = IDLE$}{
            $done \gets 0$\;
        }
        \uElseIf{$current\_state = INIT$}{
            $round \gets 0$\;
            $state \gets data\_in \oplus round\_keys[0]$\;
        }
        \uElseIf{$current\_state = ROUND$}{
            $state \gets mix\_columns\_out \oplus round\_keys[round + 1]$\;
            $round \gets round + 1$\;
        }
        \uElseIf{$current\_state = FINAL$}{
            $state \gets shift\_rows\_out \oplus round\_keys[10]$\;
            $data\_out \gets state$\;
            $done \gets 1$\;
        }
    }

    \tcp{Apply AES transformations}
    $sub\_bytes\_out \gets SubBytes(state)$\;
    $shift\_rows\_out \gets ShiftRows(sub\_bytes\_out)$\;
    $mix\_columns\_out \gets MixColumns(shift\_rows\_out)$\;

    \Return{$done$, $data\_out$}
}

\end{algorithm}

\subsection{AES\_Sub\_Bytes}
This module implements the AES S-Box substitution operation. Each byte in the 128-bit block is replaced using a lookup table or logic circuit that performs the non-linear transformation. The S-Box is designed using combinational logic or ROM-based lookup depending on FPGA resources. The \texttt{AES\_Sub\_Bytes} module implements the SubBytes transformation step in AES encryption as a synchronous process controlled by a clock and reset signals. It handles input data packets, validates them, and produces corresponding outputs based on the packet type.

\begin{itemize}
    \item \textbf{Inputs:}
    \begin{itemize}
        \item \texttt{clock}: The system clock signal, driving all sequential operations.
        \item \texttt{reset}: Active-high synchronous reset signal to initialize internal registers.
        \item \texttt{input\_valid}: A flag indicating when the input data is valid.
        \item \texttt{packet\_type}: A signal identifying the type of input packet.
        \item \texttt{input\_data}: The data input to be processed by the SubBytes operation.
    \end{itemize}
    \item \textbf{Outputs:}
    \begin{itemize}
        \item \texttt{output\_valid}: Flag indicating when the output data is valid.
        \item \texttt{output\_data}: The processed output data after the SubBytes step.
    \end{itemize}
\end{itemize}




The module operates as follows:

\begin{itemize}
    \item When the \texttt{reset} signal is asserted (high), the module clears its internal temporary data register and de-asserts the \texttt{output\_valid} signal to zero, effectively resetting its internal state.
    
    \item On each rising edge of the \texttt{clock} signal, if the \texttt{input\_valid} flag is high, the module inspects the \texttt{packet\_type}:
    \begin{itemize}
        \item If the \texttt{packet\_type} equals 2 (indicating the packet is relevant for the SubBytes operation), the input data is latched into an internal temporary register, and the \texttt{output\_valid} flag is asserted to indicate the output data is now valid.
        \item Otherwise, if the packet type does not match, \texttt{output\_valid} is de-asserted, indicating no valid output data for this cycle.
    \end{itemize}
    
    \item The output data, \texttt{output\_data}, continuously reflects the value stored in the temporary register \texttt{temp\_data}.
\end{itemize}

This behavior ensures that only valid data packets of the expected type are processed and output, while other packets are ignored.

\subsection{AES\_Shift\_Rows} This module implements the cyclic left shift of the AES state rows. The \texttt{AES ShiftRows} transformation is a permutation step applied to the 128-bit AES state matrix. It cyclically shifts the bytes in each row of the state by a certain offset to the left, depending on the row index.

\begin{itemize}
    \item \textbf{Input:} A 128-bit state \texttt{data\_in} represented as a 4x4 matrix of bytes.
    \item \textbf{Output:} A 128-bit state \texttt{data\_out}, also represented as a 4x4 matrix of bytes after applying the ShiftRows operation.
\end{itemize}


    
    
    
    


The transformation proceeds as follows:

\begin{itemize}
    \item The state matrix consists of 4 rows and 4 columns, where each element is a byte.
    \item For the first row (\texttt{row = 0}), the bytes remain unchanged; \texttt{data\_out[0][column]} is directly assigned from \texttt{data\_in[0][column]}.
    \item For subsequent rows (\texttt{row = 1, 2, 3}), each byte is shifted cyclically to the left by an amount equal to the row index:\[
\begin{aligned}
\texttt{data\_out[row][column]} &\leftarrow \\
\texttt{data\_in[row][$(column + row) \bmod 4$]} &
\end{aligned}
\] meaning each row shifts its bytes left by its row number, wrapping around cyclically.
\end{itemize}

\subsection{AES\_Mix\_Columns} The AES\_ Mix\_Columns transformation operates on the 128-bit input state \texttt{data\_in}, which is arranged as 4 columns of 4 bytes each. The transformation processes each column independently by applying finite field arithmetic in GF($2^8$).

Specifically, for each column $c$ (from 0 to 3), the four bytes $(s_0, s_1, s_2, s_3)$ of that column are extracted. Each output byte $(m_0, m_1, m_2, m_3)$ of the transformed column is computed as a linear combination of the input bytes using multiplication by constants 2 and 3 in GF($2^8$), where multiplication by 2 is implemented by the function \texttt{mul\_by\_2} and multiplication by 3 is performed by \texttt{mul\_by\_3}. These computations are defined as follows:\[
\begin{aligned}
m_0 &= \text{mul\_by\_2}(s_0) \oplus \text{mul\_by\_3}(s_1) \oplus s_2 \oplus s_3 \\
m_1 &= s_0 \oplus \text{mul\_by\_2}(s_1) \oplus \text{mul\_by\_3}(s_2) \oplus s_3 \\
m_2 &= s_0 \oplus s_1 \oplus \text{mul\_by\_2}(s_2) \oplus \text{mul\_by\_3}(s_3) \\
m_3 &= \text{mul\_by\_3}(s_0) \oplus s_1 \oplus s_2 \oplus \text{mul\_by\_2}(s_3)
\end{aligned}
\]

Here, multiplication by 2 (\texttt{ $mul_by_2$}) is implemented as a left shift of the byte followed by a conditional XOR with 0x1b if the most significant bit was set before the shift, to ensure reduction modulo of the AES polynomial. Multiplication by 3 (\texttt{mul\_by\_3}) is computed as the XOR of \texttt{mul\_by\_2} and the original byte. After computing $(m_0, m_1, m_2, m_3)$, the output state \texttt{data\_out} is updated by replacing column $c$ with these new bytes. This process is repeated for all four columns, resulting in the fully transformed AES state.

\subsection{Add\_Round\_Key and Key Scheduler} It performs XOR between the 128-bit state and the 128-bit round key. This operation is simple but crucial for combining the input data with the key material. Although not always implemented as a separate module, the AES\_Core contains logic to expand the 128-bit key into 11 round keys using the Rijndael key expansion algorithm \cite{jamil2004rijndael}. Each round key is used once per round.
\section{Testing and Evaluation Process}
To assess the practicality and performance of the proposed SPiMe architecture, we performed a detailed evaluation using multiple FPGA platforms. Our analysis spans hardware resource utilization, scalability, latency, and throughput, with designs synthesized and tested on AMD UltraScale and UltraScale+ devices.

\subsection{FPGA Platforms and Configuration}
We selected five FPGA platforms for evaluation: U55C, U280, VCU118, ZCU104, and ZCU106. These devices span a range from high-end data center accelerators to embedded-class SoCs. Table \ref{tab:devices} summarizes their hardware specifications, including available logic (LUTs), flip-flops (FFs), memory resources (BRAM and URAM), and DSP blocks.

To explore SPiME's scalability, we instantiated arrays with increasing numbers of parallel PiM units: 256, 512, 1024, 2048, and 4096. Each PIM unit comprises an AES-128 encryption core and a control unit (controller). The upper bound of 4096 units was selected based on routing limitations observed during placement and implementation, particularly due to the total wire count exceeding 1 million nets in the largest configurations.

\begin{table}[htbp]
\caption{Comparison of FPGA Devices}
\label{tab:devices}
\centering
\footnotesize
\resizebox{\linewidth}{!}{%
\begin{tabular}{lllllllll}
\toprule
\textbf{Device} & \textbf{Part} & \textbf{LUTs (K)} & \textbf{FFs (K)} & \textbf{BRAM} & \textbf{URAM} & \textbf{DSPs} \\
\midrule
U55C   & xcu55c-fsvh2892-2L-e & 1304 & 2607 & 2016 & 960 & 9024 \\
U280  & xcu280-fsvh2892-2L-e & 1304 & 2607 & 2016 & 960 & 9024 \\
VCU118 & xcvu9p-flga2104-2L-e & 1182 & 2364 & 2160 & 960 & 6840 \\
ZCU104 & xczu7ev-ffvc1156-2-e & 230  & 460  & 312  & 96  & 1728 \\
ZCU106 & xczu7ev-ffvc1156-2-e & 230  & 460  & 312  & 96  & 1728 \\
\bottomrule
\end{tabular}%
}
\end{table}

\subsection{Hardware Resource Utilization}
\begin{figure}[htp]
\centering
\includegraphics[width=\linewidth]{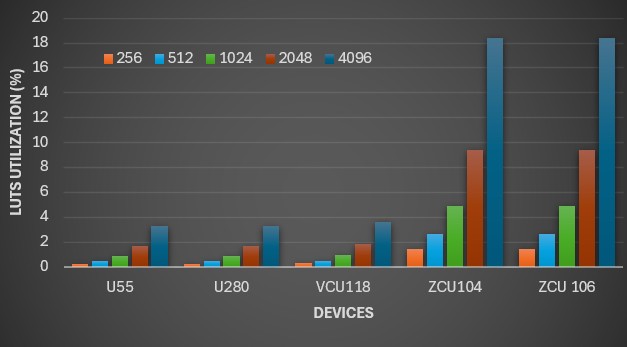} 
\caption{The LUTs utilization percentage versus number of PiMs (256-4096) on different FPGA devices.}
\label{fig:luts}
\end{figure}

Figure \ref{fig:luts} illustrates the LUT utilization across different FPGA platforms as the number of PIM units increases. On high-end devices like U55C, U280, and VCU118, even the 4096-PIM configuration consumed only around 3.65\% of the available LUTs. In contrast, smaller devices such as ZCU104 and ZCU106 reported LUT utilization near 18.44\% for the same configuration, highlighting the impact of limited logic capacity.

\begin{figure}[htp]
\centering
\includegraphics[width=\linewidth]{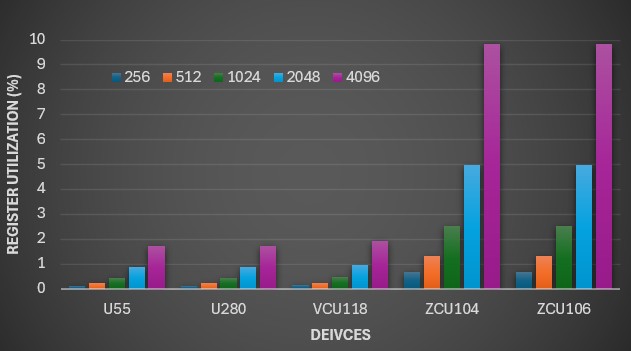} 
\caption{The register utilization percentage versus number of PiMs (256-4096) on different FPGA devices.}
\label{fig:regs}
\end{figure}

Figure \ref{fig:regs} presents the flip-flop (register) usage. As with LUTs, register consumption scales linearly with the number of instantiated PiM units. Larger FPGAs maintained sub-2\% FF utilization for 4096 PiMs, while ZCU-series devices reached nearly 10\%. These results confirm the modularity and efficiency of the SPiME design, indicating that the per-PiM overhead in both logic and control resources remains consistent across platforms.

\subsection{Latency Analysis}
Latency was calculated as a function of the number of cycles per AES operation and the operating frequency (Fmax). Each AES operation requires a fixed 11-cycle sequence: 1 cycle for initiation, 9 for AES rounds, and 1 for finalization. The latency in microseconds is given by Equation \ref{latency}:

\begin{equation}
\label{latency}
\text{Latency}~(\mu s) = 1000 \times \left( \frac{\text{Cycles to execute one task}}{f_{\text{max}}} \right)
\end{equation}

\begin{figure}[ht]
\centering
\includegraphics[width=\linewidth]{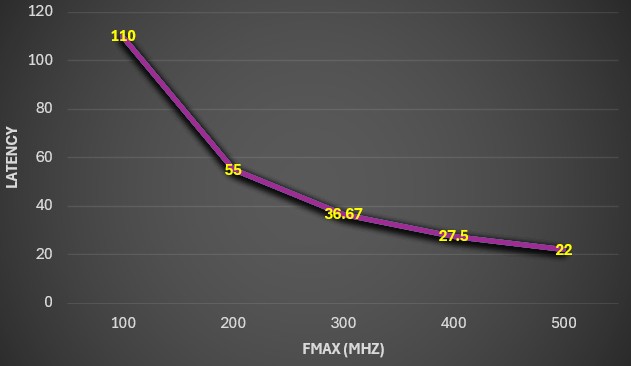} 
\caption{SPiME latency varies based on the Fmax while same for all arrays of NUM\_PIMs.}
\label{fig:latency}
\end{figure}

Figure \ref{fig:latency} shows how latency inversely scales with clock frequency. At 100 MHz, latency is approximately 0.11 $\mu$ s, reducing to ~0.036 $\mu$ s and ~0.022 $\mu$ s at 300 MHz and 500 MHz, respectively. This behavior confirms SPiMe's suitability for real-time applications, as its latency remains both low and predictable.
								
\subsection{Throughput Evaluation}
Throughput was measured as the amount of encrypted data (in bits) divided by latency. The throughput in Gbps is given by Equation \ref{throughput}:

\begin{equation}
\label{throughput}
\text{Throughput}~(\text{Gbps}) = \left( \frac{\text{Block size}}{\text{Latency}~(\mu s)} \right) \div 10^6
\end{equation}
								
\begin{figure}[htp]
\centering
\includegraphics[width=\linewidth]{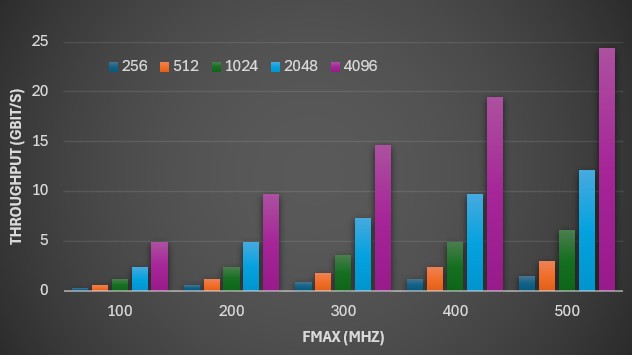} 
\caption{Throughput of 1024 block size with the NUM\_PIMs varies from 1K to 4K.}
\label{fig:throughput1K}
\end{figure}

Figure \ref{fig:throughput1K} shows the performance as the number of PIM units increases, using a fixed block size of 1024 bits. Throughput scales nearly linearly with PIM count, reaching over 23 Gbps at 500 MHz with 4096 PIMs.

\begin{figure}[h]
\centering
\includegraphics[width=\linewidth]{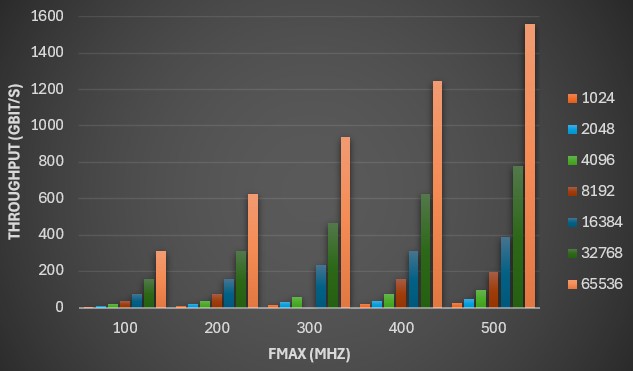} 
\caption{Throughput varies based on the block size while maximum frequency and num\_pims varies.}
\label{fig:throughput}
\end{figure}

Figure \ref{fig:throughput} shows the effect of increasing the block sizes (1K, 4K, 16K, 64K) at a fixed frequency and PIM count. Larger blocks amortize control and I/O overhead, yielding significantly higher throughput. The design performs best in batch or buffered processing scenarios, which are common in secure IoT edge devices.

The results show that SPiMe efficiently scales up to 4096 PIMs with minimal LUT and FF overhead. The latency is deterministic and low, ranging from 0.022 $\mu$ s to 0.11 $\mu$ s depending on frequency. Throughput exceeds 25 Gbps in optimal configurations, with consistent linear scaling. The architecture remains portable across large and small FPGA platforms with proportional resource usage.

\section{Conclusion and Future Work}    

This work presented SPiME, a scalable, lightweight AES-based secure memory core tailored for FPGA-based Processor-in-Memory (PiM) architectures. Implemented entirely in Verilog, SPiME integrates a PIM controller with an AES-128 core, enabling parallel encryption with minimal control overhead.

Hardware Efficiency and Scalability:
SPiME exhibits excellent scalability across a range of FPGA platforms. On high-end devices like the U55C and VCU118, we instantiated up to 4096 parallel PiM units with under 4\% resource utilization. On smaller platforms (ZCU104/106), SPiME scaled down effectively, maintaining low per-unit overhead. The design’s modularity ensures adaptability from embedded systems to datacenter-class accelerators.

Latency and Throughput:
Each AES encryption completes in a fixed 11-cycle sequence, resulting in predictable, constant-time performance. At 500 MHz, latency drops to 0.022 µs. Throughput scales linearly with the number of units and block size, reaching over 25 Gbps at peak with 4K units. Larger blocks significantly improve throughput efficiency by amortizing control overhead, making SPiME suitable for secure streaming and batched workloads like video analytics and edge AI inference.

Robustness:
No architectural bottlenecks were observed across scale tests. FSM-based control and pipelining enabled smooth operation, with routing congestion being the only limitation at extreme scales—an issue addressable in ASIC flows or future FPGAs with improved routing fabrics.

Future work includes full system integration with memory and CPU coordination, dynamic workload adaptation, real hardware benchmarking, and side-channel security validation. Additionally, SPiME will be extended with high-level software APIs to support broader adoption in edge and cloud-based secure computing environments.

To conclude, SPiME is, to our knowledge, the first FPGA-compatible, parameterizable PiM-based encryption core that supports scalable parallel AES processing. Its modular design, predictable performance, and low overhead make it a strong candidate for secure, high-throughput processing in both edge and cloud environments.
 
\bibliographystyle{unsrt}
\bibliography{conference_101719}

\begin{thebibliography}{10}

\bibitem{yang2019processing}
Xu~Yang, Yumin Hou, and Hu~He.
\newblock A processing-in-memory architecture programming paradigm for wireless internet-of-things applications.
\newblock {\em Sensors}, 19(1):140, 2019.

\bibitem{karakchi2023napoly}
Rasha Karakchi and Jason~D. Bakos.
\newblock Napoly: A non-deterministic automata processor overlay.
\newblock {\em ACM Transactions on Reconfigurable Technology and Systems}, 16:1--25, 2023.

\bibitem{karakchi2017dynamically}
Rasha Karakchi, Lothrop~O. Richards, and Jason~D. Bakos.
\newblock A dynamically reconfigurable automata processor overlay.
\newblock In {\em 2017 International Conference on ReConFigurable Computing and FPGAs (ReConFig)}, pages 1--8, 2017.

\bibitem{karakchi2019overlay}
Rasha Karakchi, Charles Daniels, and Jason Bakos.
\newblock An overlay architecture for pattern matching.
\newblock In {\em 2019 IEEE 30th International Conference on Application-specific Systems, Architectures and Processors (ASAP)}, volume 2160-052X, pages 165--172, 2019.

\bibitem{karbowniczak2025optimizing}
Ryan Karbowniczak and Rasha Karakchi.
\newblock Optimizing sequence alignment with scored nfas.
\newblock {\em arXiv preprint arXiv:2501.02162}, 2025.

\bibitem{karakchi2016hls}
Rasha Karakchi, Jordan~A. Bradshaw, and Jason~D. Bakos.
\newblock High-level synthesis of a genomic database search engine.
\newblock In {\em 2016 International Conference on ReConFigurable Computing and FPGAs (ReConFig)}, pages 1--6, 2016.

\bibitem{karakchi2024developing}
Rasha Karakchi and Ryan Karbowniczak.
\newblock Developing a self-explanatory transformer.
\newblock In {\em 2024 IEEE/ACM Symposium on Edge Computing (SEC)}, pages 523--525. IEEE, 2024.

\bibitem{jarvinen2008fully}
Kimmo Järvinen, Matti Tommiska, and Jouni Skyttä.
\newblock A fully pipelined memoryless 17.8 gbps aes-128 encryptor.
\newblock In {\em Field Programmable Logic and Applications (FPL)}, pages 147--152. IEEE, 2008.

\bibitem{xu2023pim}
Lei Xu, Hao Wang, and Yong Chen.
\newblock A processing-in-memory aes implementation in dram for secure and efficient data encryption.
\newblock {\em IEEE Transactions on Computers}, 2023.
\newblock Early Access.

\bibitem{zhang2021energy}
Jun Zhang, Yu~Liu, and Jie Han.
\newblock An energy-efficient aes implementation using approximate logic synthesis.
\newblock {\em Integration, the VLSI Journal}, 75:85--94, 2021.

\bibitem{mcevoy2006compact}
Robert McEvoy, Conor Murphy, Máire McLoone, and William Marnane.
\newblock A compact fpga-based architecture for aes encryption.
\newblock {\em IEEE Transactions on Very Large Scale Integration (VLSI) Systems}, 14(7):693--701, 2006.

\bibitem{wang2021high}
Qiang Wang, Li~Zhang, and Yifan Zhao.
\newblock High-performance aes-gcm design for 5g security on fpga.
\newblock {\em ACM Transactions on Embedded Computing Systems (TECS)}, 20(5s):1--18, 2021.

\bibitem{tiri2003securaes}
Kris Tiri and Ingrid Verbauwhede.
\newblock Securing encryption algorithms against dpa at the logic level: Next generation smart card technology.
\newblock In {\em Cryptographic Hardware and Embedded Systems (CHES)}. Springer, 2003.

\bibitem{chodowiec2003very}
Pawe{\l} Chodowiec and Kris Gaj.
\newblock Very compact fpga implementation of the aes algorithm.
\newblock In {\em International workshop on cryptographic hardware and embedded systems}, pages 319--333. Springer, 2003.

\bibitem{deshpande2009fpga}
Ashwini~M Deshpande, Mangesh~S Deshpande, and Devendra~N Kayatanavar.
\newblock Fpga implementation of aes encryption and decryption.
\newblock In {\em 2009 international conference on control, automation, communication and energy conservation}, pages 1--6. IEEE, 2009.

\bibitem{priya2022fpga}
S~Sridevi~Sathya Priya, P~Karthigaikumar, and Narayana~Ravi Teja.
\newblock Fpga implementation of aes algorithm for high speed applications.
\newblock {\em Analog integrated circuits and signal processing}, pages 1--11, 2022.

\bibitem{zambreno2004exploring}
Joseph Zambreno, David Nguyen, and Alok Choudhary.
\newblock Exploring area/delay tradeoffs in an aes fpga implementation.
\newblock In {\em International Conference on Field Programmable Logic and Applications}, pages 575--585. Springer, 2004.

\bibitem{chodowiec2002asic}
Piotr Chodowiec and Krzysztof Gaj.
\newblock Asic implementation of the aes rijndael algorithm.
\newblock In {\em International Conference on Field Programmable Logic and Applications}, pages 160--171. Springer, 2002.

\bibitem{good2005aes}
Tim Good and Mohammed Benaissa.
\newblock Aes on fpga from the fastest to the smallest.
\newblock In {\em Cryptographic Hardware and Embedded Systems--CHES 2005: 7th International Workshop, Edinburgh, UK, August 29--September 1, 2005. Proceedings 7}, pages 427--440. Springer, 2005.

\bibitem{zodpe2020efficient}
Harshali Zodpe and Ashok Sapkal.
\newblock An efficient aes implementation using fpga with enhanced security features.
\newblock {\em Journal of King Saud University-Engineering Sciences}, 32(2):115--122, 2020.

\bibitem{hasija2023survey}
Taniya Hasija, Amanpreet Kaur, KR~Ramkumar, Shagun Sharma, Sudesh Mittal, and Bhupendra Singh.
\newblock A survey on performance analysis of different architectures of aes algorithm on fpga.
\newblock {\em Modern Electronics Devices and Communication Systems: Select Proceedings of MEDCOM 2021}, pages 39--54, 2023.

\bibitem{deshpande2014efficient}
Hrushikesh~S Deshpande, Kailash~J Karande, and Altaaf~O Mulani.
\newblock Efficient implementation of aes algorithm on fpga.
\newblock In {\em 2014 International Conference on Communication and Signal Processing}, pages 1895--1899. IEEE, 2014.

\bibitem{borkar2011fpga}
Atul~M Borkar, RV~Kshirsagar, and MV~Vyawahare.
\newblock Fpga implementation of aes algorithm.
\newblock In {\em 2011 3rd International Conference on Electronics Computer Technology}, volume~3, pages 401--405. IEEE, 2011.

\bibitem{farooq2017comparative}
Umer Farooq and M~Faisal Aslam.
\newblock Comparative analysis of different aes implementation techniques for efficient resource usage and better performance of an fpga.
\newblock {\em Journal of King Saud University-Computer and Information Sciences}, 29(3):295--302, 2017.

\bibitem{zhang2018optimization}
Xiwei Zhang, Meng Li, and Jing Hu.
\newblock Optimization and implementation of aes algorithm based on fpga.
\newblock In {\em 2018 IEEE 4th International Conference on Computer and Communications (ICCC)}, pages 2704--2709. IEEE, 2018.

\bibitem{deshpande2015area}
Hrushikesh~S Deshpande, Kailash~J Karande, and Altaaf~O Mulani.
\newblock Area optimized implementation of aes algorithm on fpga.
\newblock In {\em 2015 International Conference on Communications and Signal Processing (ICCSP)}, pages 0010--0014. IEEE, 2015.

\bibitem{sunil2020implementation}
Joseph Sunil, HS~Suhas, BK~Sumanth, and S~Santhameena.
\newblock Implementation of aes algorithm on fpga and on software.
\newblock In {\em 2020 IEEE International Conference for Innovation in Technology (INOCON)}, pages 1--4. IEEE, 2020.

\bibitem{satoh2001compact}
Atsushi Satoh, Shuji Morioka, Kohji Takano, and Sumio Munetoh.
\newblock A compact {Rijndael} hardware architecture with {S-box} optimization.
\newblock In {\em Advances in Cryptology—ASIACRYPT 2001}, pages 239--254. Springer, 2001.

\bibitem{he2020parallel}
Xiaohui He, Bin Li, and Yong Zhang.
\newblock A parallel aes architecture for high-speed network security.
\newblock {\em IEEE Access}, 8:21725--21735, 2020.

\bibitem{chaves2006reconfigurable}
Ricardo Chaves, Georgi Kuzmanov, Stamatis Vassiliadis, and Leonel Sousa.
\newblock Reconfigurable memory based aes co-processor.
\newblock In {\em Proceedings 20th IEEE International Parallel \& Distributed Processing Symposium}, pages 8--pp. IEEE, 2006.

\bibitem{inspintronic}
Pegah Iranfar, Abdolah Amirany, and Mohammad~Hossein Moaiyeri.
\newblock Power attack-immune spintronic-based aes hardware accelerator for secure and high-performance pim architectures.
\newblock {\em IEEE Transactions on Magnetics}, 61(4):1--12, 2025.

\bibitem{liu2022aespim}
Yiding Liu, Guangyu Huang, Yuwei Zhang, Xuehai Wang, and Yu~Wang.
\newblock Enabling {PIM}-based {AES} encryption for online video streaming.
\newblock {\em Journal of Systems Architecture}, 132:102734, 2022.

\bibitem{lee2010processor}
Ruby~B Lee and Yu-Yuan Chen.
\newblock Processor accelerator for aes.
\newblock In {\em 2010 IEEE 8th Symposium on Application Specific Processors (SASP)}, pages 16--21. IEEE, 2010.

\bibitem{jamil2004rijndael}
Tariq Jamil.
\newblock The rijndael algorithm.
\newblock {\em IEEE potentials}, 23(2):36--38, 2004.

\end{thebibliography}

\end{document}